\documentclass[sigconf,nonacm]{acmart}
\AtBeginDocument{%
  }

\setcopyright{none}
\acmConference [GenAI-Comm@CSCW '26]{Broader Impacts of GenAI in Communication:Building Agendas for Research, Design, and Policymaking @ CSCW 2026} {October 10--14,
  2026}{Salt Lake City, Utah}

\begin{document}

\title{More Than Just Access: Generative AI as Communication Intermediary for Blind and Low-Vision Users}


\author{Protik Dey}
\affiliation{%
  \institution{University of Texas at San Antonio}
  \city{San Antonio}
  \country{Texas}}
\email{protik.dey@my.utsa.edu}

\author{Mohd Saifuzzaman}
\affiliation{%
  \institution{University of Texas at San Antonio}
  \city{San Antonio}
  \country{Texas}}
\email{mohd.saifuzzaman@my.utsa.edu}

\author{Taslima Akter}
\affiliation{%
  \institution{University of Texas at San Antonio}
  \city{San Antonio}
  \country{Texas}}
\email{taslima.akter@utsa.edu}

\renewcommand{\shortauthors}{Dey et al.}

\begin{abstract}
  Generative AI (GenAI) tools are increasingly woven into how blind and low-vision (BLV) people communicate, not only with digital information, but with the physical world and with other people. Tools such as ChatGPT, Google Gemini, Be My AI, and Seeing AI translate visual and textual content into accessible form, and are beginning to substitute for interpersonal requests for help, such as asking a family member to read a label or describe a scene. Drawing on semi-structured interviews with 19 BLV participants, we examine GenAI as a communication intermediary and how it succeeds and fails as an alternative for reading, describing, and even asking another person for help. We also investigated what BLV users gain and risk when these tools take over that role. We conclude with design and policy implications for GenAI systems that communicate uncertainty honestly, protect information, and support BLV users’ independence rather than substitute for it unsafely.
\end{abstract}

\ccsdesc[500]{Human-centered computing~Accessibility systems and tools}
\ccsdesc[300]{Human-centered computing~Empirical studies in accessibility}
\ccsdesc[300]{Human-centered computing~Empirical studies in HCI}
\ccsdesc[300]{Computing methodologies~Philosophical/theoretical foundations of artificial intelligence}

\keywords{Generative AI, Blind and Low-Vision, Communication, Trust, Accessibility, Privacy, Human-AI Interaction}


\maketitle

\section{Introduction}
Blind and low-vision (BLV) people are increasingly using Generative AI (GenAI) to communicate, for example reading a document, describing a scene, identifying an object, or asking a question that would once have required another person's help~\cite{adnin2024king(3)}. Tools such as ChatGPT, Be My AI, and Seeing AI translate the visual and textual world into a form BLV users can act on, letting them access information and, increasingly, reach other people through an AI intermediary rather than asking directly. Unlike rule-based or deterministic assistive technologies, GenAI systems produce probabilistic outputs that can be flexible and context-sensitive but also less predictable and sometimes incorrect~\cite{bender2021stochastic(1), amershi2019guidelines(2)}. For BLV users, who often cannot visually verify what a GenAI system tells them, this unpredictability is not simply a usability flaw, it is a communication problem, since the reliability of what is ``said'' to them cannot be checked against what is actually there.

This shift touches BLV users' communication in at least two distinct ways. First, it bears on their \emph{epistemic autonomy}: because BLV users often cannot independently verify a GenAI response, their capacity for information seeking and independent judgment depends on non-visual cues the system may or may not provide~\cite{adnin2024king(3), sakib2026explainable}. Second, it bears on their \emph{relationships}: GenAI is increasingly substituting for communication that once involved another person like asking a family member to read a label, describing a scene to a friend, or calling a sighted volunteer to confirm a document~\cite{sharma2025visualprivacy(42)}. In other words, swapping a person for GenAI doesn't just change how the exchange happens, it changes who the BLV user is trusting, what they're revealing, and who's responsible if it goes wrong. It also changes the effort and sincerity of the exchange itself. Asking GenAI can lower the social cost of a routine request, but it removes the emotional connection and accountability that came with asking a person~\cite{alam2025blind(4), akter2020privacy(48)}.

A wide body of work has studied why people come to rely on automated systems at all, pointing to reliability, transparency, accountability, and privacy as central considerations~\cite{jo2025trust(5), qian2024take(9), feng2024privacyqa(16), raji2020accountability(17)}. Most of this work, however, assumes a user who can visually inspect outputs, compare information, and independently verify system behavior. This assumption does not hold for BLV users communicating through GenAI. In this paper, we examine GenAI as a communication intermediary for BLV users, a channel that increasingly stands in for reading, describing, and asking another person for help. We ask how BLV users experience this substitution, what shapes their willingness to rely on it over a person, and what is gained and put at risk, for their independent judgment and their relationships, when an AI system takes over a communicative role people used to fill. To answer these questions, we conducted semi-structured interviews with 19 BLV users who had experience using a range of GenAI tools. Our findings show that GenAI succeeds as a communicative stand-in unevenly. It extends BLV users' independence and eases the everyday effort of asking family and friends, but only when the exchange itself is reliable, reachable, and safe to disclose to. BLV users are not passive about this shift, rather they exercise independent judgment about when to rely on GenAI and when to return to a person, reserving human help, and the sincerity and accountability that comes with it.

\section{Related Work}
GenAI tools are changing how BLV users communicate, not just what tools they use, but who they turn to for help and what they must share to get it~\cite{adnin2024king(3), alam2025blind(4)}. Prior work on trust in automated systems points to reliability, transparency, accountability, and privacy as central considerations~\cite{jo2025trust(5), liao2020questioning(6), qian2024take(9), feng2024privacyqa(16), raji2020accountability(17)}, and framework studies extend this across fairness, explainability, and governance~\cite{kaur2022trustworthy, thiebes2021trustworthy, li2023trustworthy, regona2026building}. This is especially relevant for GenAI, whose responses are fluent and confident even when wrong, making errors hard for anyone to catch~\cite{ji2023hallucination(7), susanto2025seahelm(8)}. Trust in these systems is not fixed but shaped by ongoing experience, not just a single correct or incorrect answer~\cite{mcintyre2025network, huynh2025trust(14), birhane2022values(15)}. Most of this work, however, assumes a user who can visually inspect outputs and verify them independently, an assumption that does not hold for BLV users. Accessibility research shows BLV users regularly encounter screen reader incompatibility, unlabeled interfaces, and inaccessible content~\cite{akter2025beyond(19), erdemli2025calendar(20), tang2025uncertainty(18)}, and recent work on explainability finds that BLV users specifically prefer conversational, back-and-forth exchanges that let them question and verify a response, over a static explanation delivered once~\cite{sakib2026explainable}.

BLV users already rely on GenAI for everyday communicative tasks such as reading documents, understanding images, and answering questions~\cite{adnin2024king(3), chang2025chatgpt(21), leporini2025preliminary(22), mo2025tablenarrator(28)}, but accessibility breaks down when a system offers no way to ask it to look again or gives no indication of what it actually looked at~\cite{gonzalezpenuela2024scenedesc(32), xie2025beyond(26), lin2025npaiai(27)}, and fluent-but-wrong responses can go unnoticed and erode trust over time~\cite{alam2025blind(4), wang2025anxiety(10)}. A smaller line of work looks at something more specific, BLV users turning to GenAI in situations where they would previously have asked a person, and carefully managing what they disclose to it in the process~\cite{sharma2025visualprivacy(42), tseng2024bivprivseg(44), zhang2024obfuscation(43)}. This mirrors long-standing findings on camera-based assistive tools, where BLV users weigh the same disclosure decision with a human helper or bystander in place of an AI, and where privacy harms such as misrepresentation and unfair treatment shape whether the helper, human or AI, is trusted at all~\cite{akter2020privacy(29), akter2020privacy(48), akter2020privacycompanion(46), akter2022shared(49)}. This substitution is not automatic; BLV users still reserve high-stakes tasks for a human they trust, and judge how much to rely on a system by how clearly it communicates its own confidence, using GenAI mainly for the smaller, more frequent requests that used to fall on people close to them~\cite{tang2025uncertainty(18), akter2024confidence(47)}. What remains underexamined is how BLV users decide, case by case, when an AI is a good enough stand-in for a person, and what they give up, in accountability, sincerity, and disclosure, when it takes that person's place.

\section{Methodology}
We conducted semi-structured interviews with 19 BLV participants with prior experience using GenAI tools, including general-purpose systems such as ChatGPT, Google Gemini, Meta AI, Perplexity, Claude, and Microsoft Copilot, as well as GenAI-powered assistive technologies such as Be My AI and Seeing AI. Recruitment was conducted primarily through the National Federation of the Blind, with a smaller number of participants recruited via snowball sampling. Interested individuals completed an online recruitment form, which also served as a screening form to assess eligibility. Eligibility criteria included: (1)~residing in the United States, (2)~age 18 years or older, (3)~identify as blind or low-vision, (4)~primarily use a screen reader for accessing digital content (instead of magnification), (5)~speak English, and (6)~have experience using GenAI tools. Two researchers screened respondents who completed the form and then contacted each eligible participant via email to schedule interviews. From a total of 33 respondents, 19 participants were selected based on their level of experience and frequency of GenAI use.

Most participants were moderate to frequent users of GenAI tools, with seven reporting use several times a day; one participant had comparatively less experience, having begun using ChatGPT approximately six months prior to the interview. All participants used GenAI-powered assistive technologies such as Be My AI or Seeing AI, and several also reported using Aira for visual assistance. To access these GenAI tools, all participants relied on screen readers, including JAWS or NVDA, and several also used mobile accessibility features such as VoiceOver or Siri. Of the 19 participants, 13 (68.4\%) identified as female and 6 (31.6\%) as male. Participants ranged across age groups, with the largest clusters between 30–39 and 40–49 years old, and educational attainment ranged from high school to doctoral degrees.

The interviews were conducted remotely via Zoom, following approval from our university's Institutional Review Board (IRB). Participants were provided with an informed consent form and a demographic questionnaire once the interview was scheduled. At the beginning of each session, the interviewer introduced the study goals and interview process and obtained participants' verbal consent to proceed.

The interview protocol was designed to investigate participants' experiences using GenAI tools and what shaped their trust in relying on GenAI to communicate with information, with their surroundings, and in place of another person. In the first part of the interview, participants were asked to describe their current interactions with GenAI systems, including the communicative tasks they used these tools for, their usage patterns, and challenges encountered. The second part explored participants' perceptions of trust and reliance through discussion of specific situations including everyday information-seeking, image and document interpretation, and moments where they chose GenAI over asking a person for help, examining how accessibility, privacy, and reliability shaped their confidence in each case. In the final part, participants were invited to share recommendations for developers and researchers on designing GenAI systems that communicate more accessibly and confidently with BLV users.

The interviews typically lasted between 45 and 60 minutes, with some extending beyond an hour to allow participants to elaborate. All interviews were video-recorded with participants' permission and later transcribed for analysis. Participants received a US\$30 Amazon gift card as compensation for their participation.

For data analysis, two authors independently engaged in the coding process using a reflexive thematic analysis approach~\cite{braun2006using}. Inter-rater reliability was not calculated, as such metrics are not considered suitable for interview-based data where codes are not treated as fixed~\cite{mcdonald2019reliability, o2020intercoder}. Instead, the authors met weekly to jointly code an initial set of interviews, discuss emerging codes, resolve differences, and refine a shared codebook. After reaching thematic saturation, the remaining interviews were divided between the two authors, who continued to meet weekly to review new codes and update shared interpretations, ultimately surfacing themes shaping whether participants trusted GenAI enough to rely on it as a communicative stand-in for a person.

\section{Findings}
Across interviews, participants described GenAI less as a single-purpose tool and more as a channel through which they accessed the visual world, digital content, and increasingly other people. Whether that channel could be trusted depended on how well it worked, whether participants could reach it at all, and what they had to give up to use it.

\subsection{Reading the World Aloud: GenAI as a Channel to Visual and Textual Information}
The most consistent use case of GenAI was translating visual or textual information into a non-visual form: describing images, reading documents, identifying objects etc. Because participants generally could not independently verify these translations, they treated the channel with calibrated skepticism rather than full acceptance. P11 explained: 

\begin{quote}
    \textit{``When I get some information, and when I read through it, you have to be wise, right? Like, do you want to really rely on the information that's being provided to you? So, if the information I'm using for my work, I read it, but I also verify and use my own judgment whether to use it for my work or not. Because sometimes, there could be a hallucination.''}
\end{quote}

Verification was easiest when participants already had background knowledge of the content being communicated to them. P15 described catching an error in a measurement task specifically because they had independent domain knowledge, not because the system flagged uncertainty. This dependence on prior knowledge is a direct constraint on epistemic autonomy: the moments when independent judgment matters most are exactly the moments when a BLV user has the least independent basis on which to exercise it. Also the channel is least trustworthy exactly when there is no other way to cross-check what they are being told, which is often the situation in which they turned to GenAI in the first place.

\subsection{From Asking a Person to Asking a Machine}
Participants described a shift from asking a person to asking GenAI tools, particularly for tasks that once required friends, family members, or bystanders. They explicitly tied this shift with independence and reduced social burden. They are also informed about the privacy and related concerns that come with using these tools for day to day tasks. P19 described this trade-off directly: \textit{``I guess in the beginning I was more wary of using them, like, taking a picture of a card or something. But I don't know if we just get used to it and just kind of look the other way, but I use them more. Not that I'm not concerned, but, like, I don't know, convenience kind of wins.''}

Participants also mentioned GenAI supporting continued connection with family and social networks by reducing their dependence on others for routine tasks, freeing interpersonal requests for situations that genuinely needed a human. This substitution trades the effort of a request against the sincerity and accountability a human respondent could offer, and participants drew that line deliberately rather than by default. They distinguished between low-stakes communicative tasks (e.g., drafting emails, making presentations) where GenAI was an acceptable replacement for a person, and high-stakes tasks (e.g., financial, medical, navigation-critical information) where they preferred a human even if it meant asking for help. P15 shared using GenAI output as a communicative starting point rather than a final answer, \textit{``Like you can use whatever that response is as a structured starting point. Like it gives you at least to have a structure like an outline and then you can go ahead and just start looking.''}

\subsection{No Accessibility, No Communication}
Regardless of how well a GenAI tool performed or how carefully it handled personal information, participants could not use it to communicate at all if the interface itself was inaccessible. Unlabeled prompt fields, unclear buttons, and broken copy-and-paste functionality directly blocked communicative exchanges before anything else about the tool could even matter. P11 said:

\begin{quote}
    \textit{``If I'll ask them a question, it will give me the answer, and JAWS will read it, but I also want to copy that answer, and I do not know where that answer is. I'm not even able to copy the answer.''}
\end{quote}

This positions accessibility not only as a factor among several, but as a threshold condition for communication itself. An inaccessible interface does not produce a lower-quality conversation, it produces no conversation at all. Participants also reported that accessibility had significantly improved over time. They also mentioned that this trajectory of improvement increased their willingness to treat the remaining barriers as temporary rather than disqualifying.

\subsection{What the Machine Knows to Help}
Because GenAI was often standing in for a person, participants weighed privacy the way one might weigh what to disclose to a stranger versus a trusted intermediary. Sharing a photo of a document, a medical form, or one's surroundings to receive a description or reading carries different stakes than doing the same with a family member. Participants described this as a conscious, negotiated trade-off rather than indifference as  convenience and independence often outweighed privacy hesitation, particularly when the alternative was depending on another person. Ownership of the underlying company shaped this perspective. Some participants avoided tools from companies they distrusted, while others prioritized accessibility and usefulness over corporate reputation. \textit{``I trust those less than Be My Eyes or even Microsoft... their products are just so useful and accessible that it doesn’t bother me as much.''}, said P17. Personalization produced a similar tension. Participants valued GenAI tools ``knowing'' their needs as BLV users, but also worried about how much the system retained about them over time.

\section{The Double-Edged Sword: Gains and Risk When GenAI Speaks for BLV Users}
Framing GenAI tools as a communication stand-in, rather than a general purpose tool, surfaces benefits and risks that a narrower usability lens tends to miss. On the benefit side, participants described gains that were social as much as technical such as fewer routine requests to family and friends, more privacy around personal matters they might not want to disclose to a person at all, and a stronger day-to-day sense of independence. These gains largely extended participants' existing relationships rather than replacing them. People still reserved interpersonal requests for situations that genuinely needed human assistance, and used GenAI to absorb the smaller, more frequent asks that used to fall on the people closest to them.

The risks follow directly from GenAI occupying a role a person used to hold. When a family member misreads a label or misdescribes a scene, the error is usually visible, correctable, and socially accountable. The person can be asked to look again, or the person may express uncertainty about the reliability of the information. When GenAI makes the same kind of error, it can be delivered with the same confidence as a correct answer. It is often harder to trace back to a cause, especially in higher stakes exchanges involving financial documents, or medical instructions. Participants managed this risk by reserving GenAI tools for low-stakes communication and returning to a person for anything consequential. But this strategy places the burden of identifying failure on the BLV user rather than on the system, and assumes the user can reliably tell in advance which exchange will turn out to be high-stakes.

A second risk concerns disclosure without full awareness of its consequences. Communicating through GenAI tools often means sharing images, documents, or spoken information that a BLV user might never disclose to a stranger, but is asked to disclose to a system whose data practices are difficult to inspect. Participants weighed this cost consciously, but not always with complete information, and this blind spot was not limited to any one group. As P8 observed, reflecting on both younger and older users: \textit{``There's gotta be some digital literacy. And that's one of the things that is a problem with my mom's generation. And, I'm not even gonna say my mom's generation. It's crazy how I guess Gen Z I work with. They grew up with it, so they don't much think about it either, you know, so there's just not that kind of awareness.''}

\section{Design Implications and Policy Recommendations}

\textbf{Communicate Failure and Uncertainty Explicitly.} Sighted users can easily infer a stalled, failed or incomplete response through visual cues, but BLV users often cannot or face difficulties. So GenAI system should treat uncertainty and failure as communicative events requiring explicit non-visual signaling. For example, the tools can announce when the process is still running or it has failed. It can also do the same if it has low confidence in the information provided. This lowers the risk that an unflagged error is mistaken for a trustworthy response giving BLV users the same failure cues a human being would provide.

\textbf{Make Accessibility a Persistent, System-Wide State.} An inaccessible interface does not just create friction, it risks shutting down the communicative exchange entirely. To reduce this risk, accessibility should be an integral part of any GenAI system development, not just an afterthought. 

\textbf{Support Task-Sensitive Communication Modes.} The risk of a miscommunication carries different consequences depending on what is at stake. Participants trusted GenAI tools differently when it was replacing a low-stakes convenience versus a high-stakes task. Systems should let users, or context, select a more conservative, uncertainty-forward mode for high-stakes communicative tasks like medical, financial, or navigation-related exchanges, versus a more exploratory mode for casual use. This reduces the risk that trust earned in low-stakes exchanges gets overgeneralized to consequential ones, a risk that our findings suggest is unevenly distributed.

\textbf{Reduce Disclosure Risk Through Accessible and Early Communication.} BLV users often disclose images, documents, and voice recordings that a sighted user might handle differently, or might not share with an unfamiliar party at all. Systems should therefore communicate, in explicit and screen-reader-accessible terms, what happens to that information (retention, reuse, third-party sharing) at the point of disclosure itself, rather than in a separate privacy policy that is rarely read and often inaccessible. Because our findings show that awareness of this risk is uneven across age groups, this communication should not assume a baseline level of digital literacy. It should be legible on first encounter, before the disclosure is made rather than after.

\textbf{Establish Baseline and Cross-Platform Protection Policy.} Participants did not want to bear the risk of individually vetting every company's privacy and accessibility practices before trusting it to carry a communicative exchange. This points toward a policy need for standardized, enforceable baseline protections which will cover data handling and accessibility conformance. This policy should be applied across GenAI products, shifting the burden of managing communication risk away from individual BLV users. Such a baseline would be especially valuable given that our findings show ownership and corporate reputation already function as an imperfect, unevenly applied proxy for trust. Some participants avoided companies they distrusted, while others set that concern aside when a tool was the only accessible option available to them.

\section{Conclusion}
This paper examined how BLV users communicate through GenAI tools when they stand in for reading, describing, and asking another person for help. Our findings show that this substitution is not automatic. BLV users exercise independent judgment about when to rely on GenAI and when a human is still needed, distinguishing low-stakes tasks from high-stakes ones and weighing reliability, accessibility, and disclosure with every exchange. These results point to the limits of treating GenAI as a general-purpose tool and underscore the need to design it explicitly as a communication partner, particularly as the cost of poor accessibility, opaque disclosure, and silent failure rises with GenAI's growing role in daily communication. Supporting BLV communication through GenAI is not a narrow accessibility problem. It is a test of whether AI systems can earn a place in relationships that used to belong to people.

\bibliographystyle{ACM-Reference-Format}
\bibliography{sample-base}

\end{document}